\documentclass[a4paper,11pt]{article}

\usepackage{contribution}

\newcommand{\weblink}[2][]{%
    \ifthenelse{\equal{#1}{}}%
    {\textnormal{\url{#2}}}%
    {\textnormal{\href{#2}{#1}}}%
}

\def\beq{\begin{equation}}
\def\eeq#1{\label{#1}\end{equation}}
\def\eeqn{\end{equation}}

\def\beqa{\begin{eqnarray}}
\def\eeqa#1{\label{#1}\end{eqnarray}}
\def\eeqan{\end{eqnarray}}

\let\bar=\overbar

\def\Dslash{\not{\hbox{\kern-4pt $D$}}}
\def\dslash{\not{\hbox{\kern-2pt $\del$}}}

\def\msb{{\bar{\ssstyle M \kern -1pt S}}}

\newcommand{\contribution}[7][]{%
  \clearpage
  \thispagestyle{plain}
  \ifthenelse{\equal{#1}{}}
  {\hypersetup{pdftitle={#2}}}
  {\hypersetup{pdftitle={#1}}}
  \hypersetup{pdfauthor={{#3} {#4}}}
  {\centering\normalfont\LARGE\bfseries\sffamily #2 \par\nobreak}
  \lhead{}
  \chead{%
    \textit{\footnotesize XIV International Conference on Hadron Spectroscopy
      (\weblink[\textit{hadron2011}]{http://www.hadron2011.de}), 13-17 June 2011, Munich, Germany}%
  }
  \rhead{}
  \bigskip
  \begin{center}
    {#3} {#4}\ifthenelse{\equal{#6}{}}{}{\footnote{\weblink[#6]{mailto:#6}}}
    \ifthenelse{\equal{#7}{}}{}{#7} \\
    \textit{#5}
  \end{center}
  \bigskip
}

\renewcommand{\abstract}[1]{%
  \begin{center}
    \begin{minipage}{0.85\textwidth}
      \begin{footnotesize}
        #1
      \end{footnotesize}
    \end{minipage}
  \end{center}
  \bigskip
}

\begin{document}

%
%
%
%
%
{ 

%

%
%

\contribution[]  
{Notes on New Narrow N$^*$}  
{Maxim V.}{Polyakov}  
{Institute for Theoretical Physics II ,
  Ruhr-University Bochum,
  D-44780 Bochum, GERMANY} 
  {}
  {}
%

\abstract{%
  We briefly discuss the most recent evidences for narrow nucleon excitation (N$^*$) with mass around 1680~MeV.
  The data show that the N$^*$ should have much stronger photocoupling to the neutron than to the proton. That makes it
  a good candidate for the anti-decuplet member. }
  \vspace{-0.3cm}

%

\subsubsection*{Theoretical predictions for anti-decuplet N$^*$}
\vspace{-0.3cm}

In this short contribution we discuss fresh evidences for the nucleon from the anti-decuplet \cite{dia}.
A detailed account for predictions and evidences for new narrow nucleon can be found in Ref.~\cite{acta}.
Main properties of N$^*$ from the anti-deculpet which were predicted theoretically in years 1997-2004 are the following:
\vspace{-0.2cm}
\begin{itemize}
\item Quantum numbers are $P_{11}$ ($J^P=\frac 12^+$, isospin=$\frac 12$) \cite{dia}.
\vspace{-0.2cm}
\item Narrow width of $\Gamma\le 40$~MeV \cite{dia,arndt,michal}.
\vspace{-0.2cm}
\item Mass of $M\sim 1650-1720$~MeV \cite{dia1,michal}.
\vspace{-0.2cm}
\item Strong suppression of  the proton photocopling relative to the neutron one \cite{max}.
This prediction was based on SU(3) flavour symmetry only. Therefore it can be used as a clear benchmark
for a nucleon member of the anti-decuplet.
\vspace{-0.2cm}
\item The $\pi N$ coupling  is suppressed, N$^*$  prefers to decay into $\eta N$, $K\Lambda$ and $\pi \Delta$ \cite{dia,arndt,michal}.
\end{itemize}
\vspace{-0.4cm}
\subsubsection*{N$^*$ in $\gamma n$ collisions}
\vspace{-0.4cm}
In the $\gamma n$ collisions (with non-suppressed exit channels such
as $\eta n$, $\gamma n$, $K_S\Lambda$, etc.) the signal of the
anti-decuplet nucleon should be seen as a prominent narrow
peak in the cross section \cite{max}. However, the
neutron is bound in a nucleus, hence  the narrow resonance signal is
hidden by  nuclear effects (by the Fermi motion at the first place\footnote{Observation of the neutron anomaly in the $\eta$ photoproduction off $^3$He \cite{LW} excludes other nuclear effects.}).
Four groups - GRAAL~\cite{gra0,gra1}, CBELSA/TAPS \cite{kru},
LNS~\cite{kas}, and Crystal Ball/TAPS~\cite{wert} - managed to
overcome this difficulty and reported evidence for a narrow
structure at $W\sim 1680$~MeV in the $\eta$ photoproduction on the
neutron (neutron anomaly\footnote{The name ``neutron anomaly" was introduced in Ref.~\cite{jetp} to denote
the bump in the quasi-free $\gamma n\to \eta n$ cross section around $W\sim 1680$~MeV and its apparent absence in the quasi-free $\gamma p\to \eta p$ cross section.}).

In year 2011 more results on the neutron anomaly were obtained.
In Ref.~\cite{Compton} the neutron anomaly was also observed in the
Compton scattering -- the study of quasi-free Compton scattering on the neutron
revealed  a narrow ($\Gamma=28\pm 12$~MeV) peak at $\sim 1685$~MeV with significance of $\sim4.6\sigma$.
Such peak is absent in the proton Compton scattering. 

\begin{figure}[htb]
  \begin{center}
    \includegraphics[width=0.4\textwidth]{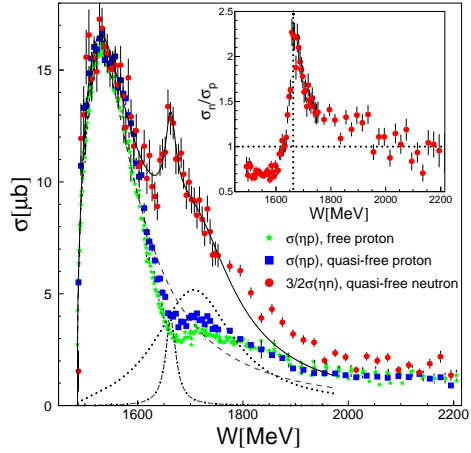}
    \caption{
    Figure from Ref.~\cite{krusche}. Total cross sections as function of final state invariant mass W with cut on spectator momentum $p_s\le 100$~MeV. (Red) dots: quasi-free neutron, (blue) squares: quasi-free proton, (green) stars: free proton data. 
     Insert: ratio of quasi-free neutron - proton data.}
    \label{fig:krusche}
  \end{center}
\end{figure}
%
In Ref.~\cite{krusche} the de-folding of the Fermi motion in quasi-free $\eta$ photoproduction off neutron
has been performed.
As a result the data exhibit  pronounced narrow ($\Gamma=25\pm12 $~MeV) peak at $W\sim 1670$~MeV
in the total cross section of $\gamma n\to \eta n$ shown in Fig.~\ref{fig:krusche}.

Looking at this figure, the first natural hypothesis is that the peak is due to
 contribution of a narrow nucleon resonance. However, due to 
the very negative attitude of the community to narrow pentaquarks (see e.g. \cite{Close}) one tries  to find another explanation for the neutron anomaly first. Detailed discussion of the ``conventional explanations" can be found in Ref.~\cite{acta}.
Some of them are refuted already by recent experimental data of Refs.~\cite{Compton,krusche}. Here we touch  presently popular 
results of Ref.~\cite{dor} only. Ref.~\cite{dor} attributes the peak in the neutron channel (see e.g. Fig.~1) to the $K Y$ threshold cusp effects. 

A dedicated experimental search of the $K Y$ threshold cusp effects was performed in Ref.~\cite{cusp}. A very small 
effect was found. Our studies (in preparation) showed that if the peak in Fig.~1 is due to the cusp effects it would imply that
$S_{11}(1650)$ resonance must have extraordinarily  large coupling to $KY$ channels, in acute disagreement with
flavour SU(3). Moreover several
questions to cusp effects  of   Ref.~\cite{dor} remain unanswered: 1) Why the neutron anomaly is absent in the pion photoproduction?
2) What is the physics reason for very fine cancelation (fine tuning) of  the $K Y$ threshold cusp effects in the proton channel? 

\begin{table}[tb]
  \begin{center}
    \begin{tabular}{lccc}  
    observable   &     extracted value &      refs. (neutron data) & refs. (proton data)  \\
    \hline
      mass (MeV)      &$1680\pm 15$  & \cite{gra0,gra1,kru,kas,wert,Compton,krusche}\cite{arndt}$^{\star)}$& \cite{acta,jetp,KPT,BG}  \cite{arndt}$^{\star)}$\\
     $\Gamma_{\rm tot}$ (MeV)  & $\le 40$     &    \cite{gra0,gra1,kru,kas,wert,Compton,krusche}\cite{arndt}$^{\star)}$& \cite{acta,jetp,KPT,BG} \cite{arndt}$^{\star)}$ \\
      $\Gamma_{\pi N}$ (MeV) &    $\le 0.5$ &      \cite{arndt}$^{\star)}$&  \cite{arndt}$^{\star)}$ \\
      $\sqrt{{\rm Br}_{\eta N}} A_{1/2}^n \ (10^{-3}\ {\rm GeV}^{-1/2})$   & 12-18     &      \cite{az,krusche} & \\
$\sqrt{{\rm Br}_{\eta N}} A_{1/2}^p \ (10^{-3}\ {\rm GeV}^{-1/2})$         & 1-3         &                                  &   \cite{acta,jetp,KPT,BG}  \\
    \end{tabular}
    \caption{{\it Our estimate} of properties of the putative narrow N$^*$ extracted from the data. 
    $^{\star)}$In Ref.~\cite{arndt} the elastic $\pi N$ scattering data were analyzed
    and the tolerance limits for N$^*$ parameters were obtained. The preferable  quantum numbers in this analysis are $P_{11}$.}
    \label{tab:blood}
  \end{center}
\end{table}
%
\vspace{-0.5cm}
\subsubsection*{N$^*$ in $\gamma p$ collisions}
\vspace{-0.4cm}
The first search of the putative anti-decuplet nucleon in $\gamma p\to
\eta p$ process was performed in Refs.~\cite{acta,jetp}. It was
found that the beam asymmetry $\Sigma$ exhibits  a sharp structure around $W\sim1685$~MeV.
That structure looks like a peak at forward angles  which develops
 into an oscillating structure at larger scattering angles. Such
a behaviour may occur by interference of a narrow resonance with a
smooth background. The observed structure was identified in
Refs.~\cite{acta,jetp} with the contribution of a resonance with
mass $M\sim 1685$~MeV, narrow width of $\Gamma\le 25$~MeV, and small
photo-coupling of $\sqrt{{\rm Br}_{\eta N}} A_{1/2}^p \sim
(1-2)\cdot 10^{-3}$~GeV$^{-1/2}$.

About an year ago the Crystal Ball Collaboration at MAMI published high
precision data on $\eta$ photoproduction on the free proton
\cite{Mainz}. The cross section was measured in fine steps in photon
energy. The measured cross section exhibits an oscillating with energy structure 
around 1690~MeV. The best fit to the data was achieved with a new version
of SAID (GE09) \cite{Mainz}. However, inspection of this fit reveals a systematic
deviation of data from the fit curves in the $1650-1730$~MeV region.
In \cite{KPT} this deviation was interpreted as
indication for a nucleon resonance with mass of $M\sim 1685$~MeV, a
narrow width of $\Gamma\leq 50 $~MeV, and a small resonance
photo-coupling  in the range of $\sqrt{{\rm Br}_{\eta N}} A_{1/2}^p
\sim (0.3-3)\cdot 10^{-3}$~GeV$^{-1/2}$. In this case no PWA of the
data was performed as needed to decide whether or not a resonance occurs
in a certain partial wave.

Such PWA was performed in Ref.~\cite{BG}.
A fit using only known  broad
resonances and standard background amplitudes can not describe
the relatively narrow oscillating structure in the cross section in the
mass region of 1660-1750~MeV.  An improved
description of the data can be reached by either assuming the
existence of a narrow resonance at a mass of about 1700~MeV with
small photo-coupling or by a threshold effect. In the latter case
the observed structure is explained by a strong (resonant or
non-resonant) $\gamma p\to\omega p$ coupling in the $S_{11}$ partial
wave. When the beam asymmetry data of Refs.~\cite{acta,jetp}
are included in the fit, the solution with a narrow
$P_{11}$ state is preferred. In that fit, mass and width of
the putative resonance converge to  $M\sim$1694~MeV and
$\Gamma\sim 40$~MeV, respectively, and the photo-coupling to
$\sqrt{{\rm Br}_{\eta N}} A_{1/2}^p \sim 2.6\cdot
10^{-3}$~GeV$^{-1/2}$.

In Table~1 we summarize   {\it our estimates} of the properties of the narrow N$^*$ which can be extracted from the present data. 
The obtained values fit neatly to the predicted properties of the anti-decuplet N$^*$. Future experiments,
especially on double polarization neutron observables, will show whether an analogous Table will appear in PDG.


%

}  


\end{document}